\documentclass[11pt]{article}

\AtBeginDocument{%
  }

\usepackage[T1]{fontenc}      
\usepackage[utf8]{inputenc}   
\usepackage{lmodern}          
\usepackage{graphicx}         

\usepackage{amsmath, amssymb} 
\usepackage{booktabs}         
\usepackage{subcaption}       
\usepackage{xcolor}           
\usepackage{siunitx}          
\usepackage{hyphenat}         
\usepackage{float}
\usepackage{algorithm, algpseudocode} 
\usepackage{natbib} 
\usepackage{listings}
\usepackage{xcolor}
\usepackage{hyperref}         

\setcitestyle{square,aysep={,}} 

\begin{document}

\title{Half Pound Filter for Real-Time Animation Blending}

\author{
Riccardo Lasagno\thanks{riccardo.lasagno@gmail.com} \\
Sumo Digital Ltd \\
Cambridge, United Kingdom
}

\date{}

\hypersetup{
    pdftitle={Half Pound Filter for Real-Time Animation Blending},
    pdfauthor={Riccardo Lasagno},
    pdfkeywords={computer graphics, animation blending, real-time rendering}
}

\maketitle

\begin{abstract}
This paper introduces the Half Pound Filter (HPF) as a modification of the 1 Euro Filter (1EF) and algorithms for automatic data-driven tuning and for filter triggering based on motion derivative boundary checks. An application of the filter is presented in the context of human animation replay for real-time simulations, where switches in animation clips cause discontinuities that must be hidden by filtering the bone trajectory without introducing noticeable artifacts. The quality of the filtering will be compared with other common animation filtering techniques using an example case drawn from the LaFAN1 dataset, showing that the resulting animation is replayed with higher fidelity by evaluating the Mean Squared Error (MSE) and Normalized Power Spectrum Similarity (NPSS) for each setup. Performances will be evaluated using both a standard predetermined trigger point and blending distance and the automatic blending trigger and recovery system.
\end{abstract}

\section{INTRODUCTION}
In real-time simulated environments, animation playback is usually controlled by a Finite State Machine (FSM) where each state plays an animation clip ~\cite{UnrealStateMachines} ~\cite{UnityAnimationStateMachine}. Transitions between states cause discontinuities that have to be masked by generating in-between poses. This is often achieved by sampling both the old state animation clip pose and the new state animation clip pose and interpolating between them for a fixed amount of time. This is effective in many scenarios, but suffers from key limitations ~\cite{BolloGDCInertialization}:
\begin{enumerate}
    \item Blending is not context-aware and keeps interpolating regardless of pose proximity
    \item Adding context adds significant complexity and performance overhead
    \item Tuning interpolation is costly and scales poorly with FSM size and complexity
    \item Blending requires to sample both animation clips
    \item In networked systems, synchronizing the FSM over clients becomes complex and prone to artifacts
\end{enumerate}
These challenges lower animation fidelity, which is problematic given the high cost and artistic intent of authored assets. To avoid reliance on FSM, data-driven techniques like Motion Matching ~\cite{Clavet2016Motion} have been proposed, which lessen the problem by providing more closely matching pose stream, but still needs smoothing ~\cite{MotionFields_10.1145/1882261.1866160}. A common workaround is to defer the blending to a post-process stage after the system determines the next animation state. The current industry standard for high-performance animation transitions is Inertialization, as presented by Bollo for Gears of War 4~\cite{BolloInertialization}. This technique achieves a performance gain by replacing traditional blending with an additive post-process that matches the new pose's position and velocity while enforcing zero jerk via a fifth-order polynomial, with the transition duration as the sole tunable parameter. While widely adopted for its efficiency and quality, this approach can introduce over-smoothing artifacts when applied to valid, high-frequency movements and over-shooting when applied to fast changing motion.
The 1EF is a simple, speed adaptive, model-free low-pass filter with a single parameter to tune, making it a suitable candidate for motion smoothing tasks ~\cite{1EF_10.1145/2207676.2208639}. HPF is derived from the 1EF, and we show that it is possible to introduce a completely automatic data-driven tuning procedure. By combining the proposed HPF filter with an automatic policy to detect and smooth discontinuities one can achieve high-fidelity animation replay without paying the cost of tuning cycles, making it scalable for real-world productions.
The results are compared using Mean Square Error and Normalized Power Spectrum Similarity ~\cite{gopalakrishnan2019neuraltemporalmodelhuman}, using a sample generated using two animation joined from the LaFAN1 dataset ~\cite{harvey2020robust}, to show that the HPF filter improve the quality of filtering without introducing meaningful frequency distortion.

\section{RELATED WORK}

\subsection{Animation Blending in Real-Time Environments}
The foundation of real-time character animation rests on the ability to transition smoothly between motion clips. Early systems established the Animation Blend Tree as the standard hierarchical framework for motion synthesis ~\cite{Perlin1995}. This approach utilizes nodes using linear interpolation to combine clips based on normalized weights. ~\cite{Rose1998} expanded this into a multi-dimensional parameter space with Verb-Adverbs, allowing for the interpolation of motions based on emotional or physical "adverbs" (e.g., walking happily vs. sadly). Motion Warping ~\cite{Witkin1995} and Inverse Kinematic are used for steady-target in-betweening, where an animation is deformed to meet a static constraint, such as reaching for a fixed handle. To avoid complex state bookkeeping during transitions, Inertialization ~\cite{BolloInertialization}, a high-performance alternative to cross-fading that decays the difference between source and target states, was introduced and widely adopted.

\subsection{Motion Matching and Data-Driven Techniques}
To manage large datasets without manual trees, Motion Graphs ~\cite{MotionGraphs_10.1145/3596711.3596788} automated the discovery of transition points. Recent research has shifted from graph structures toward Motion Matching ~\cite{Bulow2016} ~\cite{Clavet2016Motion}, a data-driven strategy that continuously selects the best frame from a large database based on the character's current state and a predicted trajectory. To solve the memory and search overhead of this method, Learned Motion Matching ~\cite{Holden2020} uses neural networks to compress the database and speed up selection. Beyond matching, generative architectures like Phase-Functioned Neural Networks ~\cite{Holden2017} and Neural Animation Layering ~\cite{Starke2021} synthesize motion by learning the underlying manifold of human movement. Recent breakthroughs have introduced Diffusion-based models like AnyTop ~\cite{Gat2025}, which enables motion synthesis across arbitrary character topologies, and Metric-Aligning Motion Matching ~\cite{Agata2025}, which optimizes transitions for physical accuracy.

\subsection{Digital filters for Motion Smoothing}
To eliminate high-frequency jitter or foot sliding, digital filters are often integrated into the animation pipeline. While the Butterworth filter is a staple for offline processing, real-time systems favor the 1€ Filter ~\cite{1EF_10.1145/2207676.2208639} due to its ability to adaptively balance jitter reduction with low latency. For robust state estimation, Kalman Filters (KF) are used to denoise joint positions by weighting kinematic predictions against observations ~\cite{Kirk2005}. Because joint rotations are non-linear, the Extended Kalman Filter (EKF) is employed to linearize these distributions ~\cite{Welch1995}~\cite{Vlasic2007}.

\section{FROM ONE EURO FILTER TO HALF POUND FILTER}
The 1EF is formulated as a digital RC low-pass filter where the cut frequency varies proportional to the magnitude of the smoothed velocity of the signal. By changing the calculation of the signal cutoff frequency to a linear interpolation from a minimum to a maximum frequency driven by the ratio of the absolute value of the time derivative of the signal with the maximum absolute value of the time derivative, we get the HPF.

\begin{algorithm}
\caption{Half Pound Filter}
\begin{algorithmic}[1] 
\State {Offline tuning parameters:} $f_{c,min}, f_{c,max}, \max{|\overset{\cdot}{x}|}$
\State {State variables:} $\hat{x}_{i-1}$
\State {Return:} $\hat{x}_i$  
\Procedure{HPF}{$x_i, \Delta t$}
    \Statex
    \If{$\mathrm{IsInvalid} \, \left(\hat{x}_{i-1} \right)$}
        \State $\hat{x}_{i-1} \gets x_i$
        \State \textbf{return} $x_i$
    \EndIf
    \Statex
    \State $\overset{\cdot}{x}_i \gets \tfrac{x_i - \hat{x}_{i-1}}{\Delta t}$
    \State $\alpha_f \gets \tfrac{|\overset{\cdot}{x}_i|}{\max{|\overset{\cdot}{x}|} }$
    \State $\alpha_f \gets \mathrm{clamp01}\, \left( \alpha_f \right)$
    \State $f_c \gets {\left(1 - \alpha_f \right)} f_{c,min} \, \text{+} \, {\alpha_f} f_{c,max} $
    \State $\alpha = \tfrac{1}{1 \, \text{+} \, \tfrac{1}{2 \pi f_{c} \Delta t}}$
    \State $\hat{x}_i \gets {\left(1 - \alpha \right)} \hat{x}_{i - 1} \, \text{+} \, {\alpha} x_i$
    \State $\hat{x}_{i-1} \gets \hat{x}_i$
    \Statex
    \State \textbf{return} $\hat{x}_i$
    \Statex
\EndProcedure
\end{algorithmic}
\end{algorithm}

\begin{figure}[t]
  \centering
  \includegraphics[width=\columnwidth]{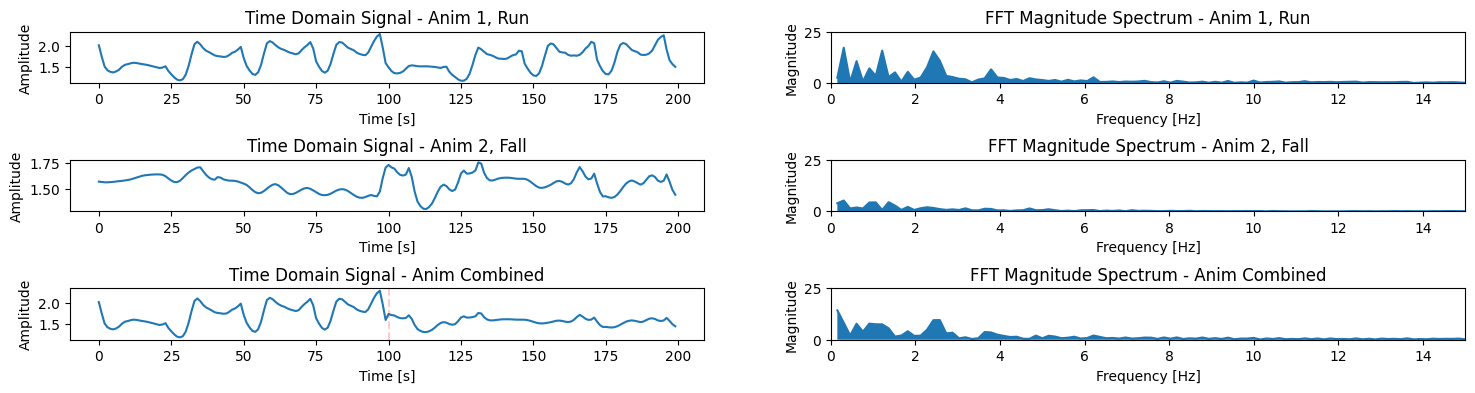}
  \hfill
  \includegraphics[width=\columnwidth]{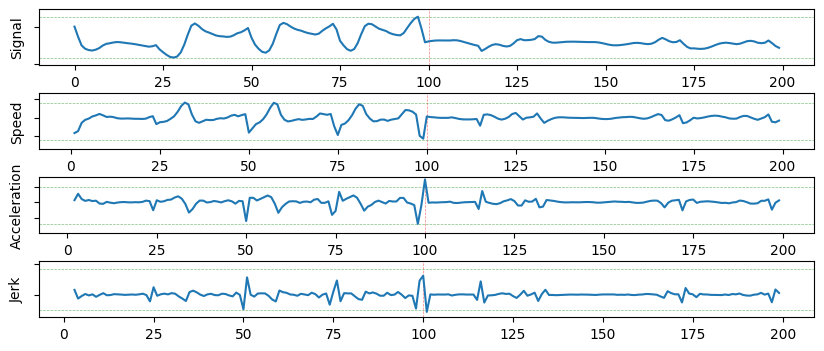}
  \caption{The generated example signal and its derivatives. The signal was created by joining a run animation and a fall animation from the LaFAN1 dataset, and the Pitch Local Euler angle for the knee is used. The horizontal dashed lines are the upper and lower boundaries, while the vertical line shows where the two animations have been joined. notice that the only places where the boundaries are violated by acceleration and jerk match the joint frame. }
  \label{fig:hpf-fix-win-comp}
\end{figure}

Note that the scope of the filter can be extended to higher derivatives by introducing additional frequency boundaries to smooth the derivatives. Since this would essentially stack two or more HPF on top of each other, these filters will be referred to as 1PF, 1.5PF, 2PF etc depending on how many successive derivatives of the signal are smoothed.

The filter acts as an interpolation, so it would never overshoot the boundaries of the animation even in limit conditions, compared to Inertialization techniques that are prone to this kind of issues since they project the previous velocity to calculate the new pose. Bollo's Inertialization overcome partially this issue by setting the projected velocity to zero if it would lead the projected value further away from the target value. 

\section{TUNING HPF FOR REAL-TIME ANIMATION BLENDING}

Character animation assets in video games are typically sampled at a frequency of 30 frames per second (fps). According to the Nyquist-Shannon sampling theorem, this sampling rate establishes a natural upper bound on the resolvable frequency content at 15 Hz. 
The majority of the power spectrum of natural human movement is known to be concentrated in the lower frequencies, typically below 5 Hz, meaning the 30 fps capture rate is generally sufficient to avoid aliasing.
We can systematically analyze the kinetic content of the animation data by scanning either the entire library of motion clips or a consolidated, representative sequence (often referred to as a "dance card" clip). This analysis allows us to precisely quantify the maximum derivatives (velocity and acceleration) and dominant frequencies present in the authored movement. Using the slope-derivative relation for band-limited signals, we can get a conservative approximation for $f_{c,min}$ as:
\[
f_{c, min} = \frac{\max{|\overset{\cdot}{x}|}}{2 \pi \max{|x|}}
\] 
We can then extract $f_{c, max}$ by finding the frequency that contains 99.99\% of the total power spectrum. As the initial frequency estimates may prove too conservative, resulting in excessive smoothing, an optional tuning gain can be applied to shift the filter's working frequencies upward. This adjustment should be implemented cautiously to ensure the system's effective frequency response remains strictly below the 15 Hz Nyquist frequency.

At the start of the smoothing process, it is preferable to use a more conservative cutoff frequency to prevent the filtered signal from converging toward the target too quickly, which might introduce excessive jerk. In later stages, the primary objective shifts to ensuring that the signal aligns closely with the target in order to avoid end-of-interval jerk. This motivates progressively increasing the frequency bounds over the course of the smoothing procedure. We refer to this strategy as Gain Blend (GB).

While using the HPF, the filtering behavior is dominated by the dynamics of the frequency shift. As a result, the method becomes less sensitive to small variations in the chosen frequency bounds, allowing the use of approximate values. In practice, for human motion data, it is reasonable to set $f_{c,min}=1$ Hz and $f_{c,max}=5$ Hz for all channels, and to linearly interpolate both values during the blend to  15 Hz. This yields a practical, ready-to-use filter that requires no tuning at all. In this paper, we refer to this configuration as GB-HPF.

\begin{figure*}[t]
  \centering
  \includegraphics[width=\textwidth]{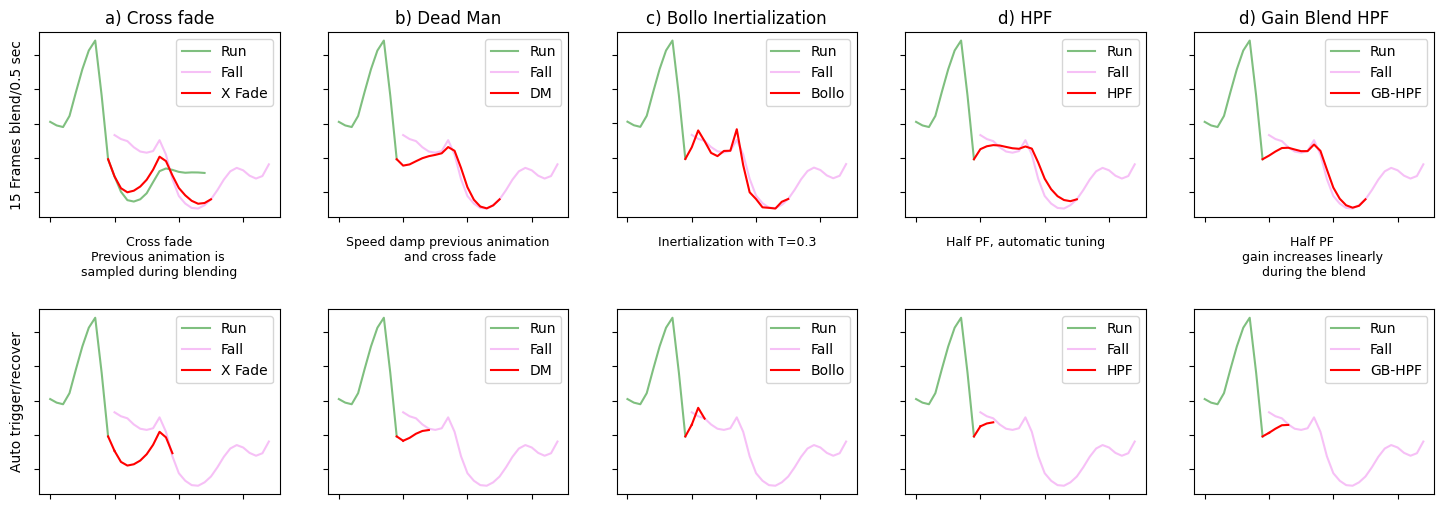}
  \caption{Overview of the different filters applied to a noisy signal. The first row shows the resulting smoothing for different techniques applied in a fixed time window, while the second row show the same filters using the automatic trigger and recovery policy.}
  \label{fig:hpf-comparison}
\end{figure*}

\begin{algorithm}[H]
\caption{Automatic Trigger and Recover with HPF}
\begin{algorithmic}[1] 
\State {State variables:} ${x}_{i-1}, \hat{x}_{i-1}, \Delta t_{i-1}, \hat{x}_{i-2}, \Delta t_{i-2},  \hat{x}_{i-3}$
\State {Return:} $\hat{x}_i$  
\Procedure{AutoHPF}{$x_i, \Delta t, active_{i-1}$}
    \Statex
    \If{$\mathrm{NotEnoughSamples} $}
        \State \Call{UpdateAutoHPFState}{$x_i, \Delta t_i$}
        \State \textbf{return} $x_i$
    \EndIf
    
    \Statex
    \State $v_0 \gets \left( x_i - \hat{x}_{i -1} \right) \text{/} \Delta t_i$
    \State $v_1 \gets \left( \hat{x}_{i-1} - \hat{x}_{i -2} \right) \text{/} \Delta t_{i-1}$
    \State $v_2 \gets \left( \hat{x}_{i-2} - \hat{x}_{i -3} \right) \text{/} \Delta t_{i-2}$
    \State $a_0 \gets \left( v_0 - v_1 \right) \text{/} \Delta t_{i}$
    \State $a_1 \gets \left( v_1 - v_2 \right) \text{/} \Delta t_{i-1}$
    \State $j \gets \left( a_0 - a_1 \right) \text{/} \Delta t_{i}$

    \Statex
    \State $active = \neg \mathrm{AreInRange} \left(x_i, v_0, a_0, j \right) $
    
    \Statex
    \If{$\neg active \land active_{i-1}$}
        \State $v_0^t  \gets \left( x_i - x_{i -1} \right) \text{/} \Delta t_i$
        \State $a_0^t  \gets \left( v_0 - v_0^t \right) \text{/} \Delta t_i$
        \State $active = active \lor \left( a_0^t  > a_{0, max}^t  \right)  \lor \left( a_0^t  < a_{0, min}^t  \right)$
    \EndIf

    \Statex
    \If{$\neg active$}
        \State \Call{UpdateAutoHPFState}{$x_i, \Delta t_i$}
        \State \textbf{return} $x_i$
    \EndIf

    \Statex
    \State $\hat{x}_i = \text{HPF} \left(x_i, \Delta t_i \right)$
    \State \Call{UpdateAutoHPFState}{$\hat{x}_i, \Delta t_i$}
    
    \State \textbf{return} $\hat{x}_i$
\EndProcedure
\Statex
\Procedure{UpdateAutoHPFState}{$x_i, \Delta t_i$}
    \State $\hat{x}_{i-3} \gets \hat{x}_{i-2}$
    \State $\hat{x}_{i-2} \gets \hat{x}_{i-1}$
    \State $\hat{x}_{i-1} \gets x_i$
    \State $\Delta t_{i-2} \gets \Delta t_{i-1}$
    \State $\Delta t_{i-1} \gets \Delta t_i$
\EndProcedure
\end{algorithmic}
\end{algorithm}

\section{AUTOMATIC TRIGGER AND RECOVERY POLICY FOR MOTION FILTERS}

The conventional approach applies a filter over a fixed time window upon every clip change. This non-context-aware policy often leads to over-smoothing artifacts (e.g., "foot skating") because the filter is triggered regardless of the transition's inherent smoothness and continues to run for the full duration, even if the filtered signal overshoots the target pose before the window expires. 
A more robust approach relies on kinematic constraint checking to detect genuine discontinuities. Since human perception is highly sensitive to sudden changes in acceleration, we define and enforce minimum and maximum limits on the signal's derivatives, up to the third derivative (jerk). Smoothing is only triggered if the new animation sample violates one of these boundaries. While the filter is active, if the resulting signal doesn't violate the derivative boundaries but its velocity and the velocity of the target are too far apart, the filter is kept active \ref{fig:hpf-fix-win-comp}. This policy ensures that filtering is applied solely when required without introducing jerk artifacts. 

\begin{table}[htbp]
  \centering
  \caption{Filter performance comparison}
  \label{tab:filter-performance}
  \begin{tabular}{lrr}
    \hline
     Name         &      MSE &    NPSS \\
    \hline
     Raw Combined &   0.0000 & 0.0445 \\
     XFade        &   0.0259 & 0.0414 \\
     DeadMan      &   0.0054 & 0.0439 \\
     Bollo        &   0.0017 & 0.0436 \\
     HPF          &   0.0032 & 0.0464 \\
     GB-HPF       &   0.0019 & 0.0446 \\
     XFade Auto   &   0.0215 & 0.0409 \\
     DeadMan Auto &   0.0035 & 0.0434 \\
     Bollo Auto   &   0.0005 & 0.0443 \\
     HPF Auto     &   0.0006 & 0.0440 \\
     GB-HPF Auto  &   0.0014 & 0.0438 \\
    \hline
    \end{tabular}
\end{table}

\section{EVALUATING RESULTS}

We will evaluate results using 2 metrics:
\begin{enumerate}
    \item \textbf{Mean Squared Error (MSE)}: Distance between the filtered animation and the target animation, averaged. We will use fixed frame length to average the results, so that they can be compared.
    \item \textbf{Normalized Power Spectrum Similarity (NPSS)}: Weighted average of the differences of the normalized power spectrum. It is meant to show how well frequency features are respected.
\end{enumerate}

To have a representative spectrum to compare, we will use the two full source animations of about 5000 frames each to get an estimate of the overall normalized power spectrum by taking the max value for each frequency in the spectrum before computing the normalized power spectrum.

Table \ref{tab:filter-performance} shows for different techniques that while the MSE is reduced by using HPF and GB-HPF, especially with the automatic trigger and recovery, their NPSS values are similar or slightly below the raw combined animation, showing that they all reasonably respect the power spectrum features extracted by the two original animations.

\section{CONCLUSIONS}

The HPF filter family enables an animation smoothing framework that supports automatic, data-driven parameter tuning while maintaining high-fidelity motion reconstruction and avoiding over-shooting. This makes it a scalable solution for animation production and mitigates the limitations associated with inertialization approaches.
\bibliographystyle{plainnat}
\bibliography{references}

\end{document}